\PassOptionsToPackage{unicode}{hyperref}
\PassOptionsToPackage{hyphens}{url}
\documentclass[
]{article}
\usepackage{amsmath,amssymb}
\usepackage{iftex}
\ifPDFTeX
  \usepackage[T1]{fontenc}
  \usepackage[utf8]{inputenc}
  \usepackage{textcomp} 
\else 
  \usepackage{unicode-math} 
  \defaultfontfeatures{Scale=MatchLowercase}
  \defaultfontfeatures[\rmfamily]{Ligatures=TeX,Scale=1}
\fi
\usepackage{lmodern}
\ifPDFTeX\else
\fi
\IfFileExists{upquote.sty}{\usepackage{upquote}}{}
\IfFileExists{microtype.sty}{
  \usepackage[]{microtype}
  \UseMicrotypeSet[protrusion]{basicmath} 
}{}
\makeatletter
\@ifundefined{KOMAClassName}{
  \IfFileExists{parskip.sty}{%
    \usepackage{parskip}
  }{
    \setlength{\parindent}{0pt}
    \setlength{\parskip}{6pt plus 2pt minus 1pt}}
}{
  \KOMAoptions{parskip=half}}
\makeatother
\usepackage{xcolor}
\usepackage[margin=1in]{geometry}
\usepackage{graphicx}
\makeatletter
\def\maxwidth{\ifdim\Gin@nat@width>\linewidth\linewidth\else\Gin@nat@width\fi}
\def\maxheight{\ifdim\Gin@nat@height>\textheight\textheight\else\Gin@nat@height\fi}
\makeatother
\setkeys{Gin}{width=\maxwidth,height=\maxheight,keepaspectratio}
\makeatletter
\def\fps@figure{htbp}
\makeatother
\NewDocumentCommand\citeproctext{}{}

\makeatletter
 \let\@cite@ofmt\@firstofone
 \def\@biblabel#1{}
 \def\@cite#1#2{{#1\if@tempswa , #2\fi}}
\makeatother
\newlength{\cslhangindent}
\newlength{\csllabelwidth}
\newenvironment{CSLReferences}[2] 
 {\begin{list}{}{%
  \setlength{\itemindent}{0pt}
  \setlength{\leftmargin}{0pt}
  \setlength{\parsep}{0pt}
  \ifodd #1
   \setlength{\leftmargin}{\cslhangindent}
   \setlength{\itemindent}{-1\cslhangindent}
  \fi
  \setlength{\itemsep}{#2\baselineskip}}}
 {\end{list}}
\usepackage{calc}

\newcommand{\CSLLeftMargin}[1]{\parbox[t]{\csllabelwidth}{\strut#1\strut}}
\newcommand{\CSLRightInline}[1]{\parbox[t]{\linewidth - \csllabelwidth}{\strut#1\strut}}

\usepackage{listings}
\usepackage{setspace}
\usepackage{pdfcomment}

\usepackage{footnote}
\makesavenoteenv{table}

\ifLuaTeX
  \usepackage{selnolig}  
\fi
\usepackage{bookmark}
\IfFileExists{xurl.sty}{\usepackage{xurl}}{} 
\hypersetup{
  pdftitle={The expected value of sample information calculations for external validation of risk prediction models},
  pdfauthor={Mohsen Sadatsafavi*; Andrew J Vickers; Tae Yoon Lee; Paul Gustafson; Laure Wynants},
  pdfkeywords={Risk prediction; Value of Information; Uncertainty;
Beysian statistics''},
  hidelinks,
  pdfcreator={LaTeX via pandoc}}

\title{The expected value of sample information calculations for
external validation of risk prediction models}
\author{Mohsen Sadatsafavi* \and Andrew J Vickers \and Tae Yoon
Lee \and Paul Gustafson \and Laure Wynants}
\date{December 5, 2024}

\begin{document}
\maketitle
\begin{abstract}
In designing external validation studies of clinical prediction models,
contemporary sample size calculation methods are based on the
frequentist inferential paradigm. One of the widely reported metrics of
model performance is net benefit (NB), and the relevance of conventional
inference around NB as a measure of clinical utility is doubtful. Value
of Information methodology quantifies the consequences of uncertainty in
terms of its impact on clinical utility of decisions. We introduce the
expected value of sample information (EVSI) for validation as the
expected gain in NB from conducting an external validation study of a
given size. We propose algorithms for EVSI computation, and in a case
study demonstrate how EVSI changes as a function of the amount of
current information and future study's sample size. Value of Information
methodology provides a decision-theoretic lens to the process of
planning a validation study of a risk prediction model and can
complement conventional methods when designing such studies.
\end{abstract}

\let\thefootnote\relax\footnotetext{From Faculty of Medicine and Pharmaceutical Sciences (MS, TYL), and Department of Statistics (PG), The University of British Columbia, Vancouver, British Columbia, Canada; Department of Epidemiology and Biostatistics, Memorial Sloan Kettering Cancer Center, New
York, New York, USA (AV); Department of Epidemiology, CAPHRI Care and Public Health Research Institute, Maastricht University, Maastricht, The Netherlands, and Department of Development and Regeneration, KU Leuven, Leuven, Belgium (LW)}
\let\thefootnote\relax\footnotetext{*Correspondence to Mohsen Sadatsafavi, 2405 Wesbrook Mall, Vancouver, BC, V6T1Z3, Canada; email: mohsen.sadatsafavi@ubc.ca}

\graphicspath{{./}}

\section{Introduction}\label{introduction}

Risk prediction models that quantify the probability of clinical events
are a key aspect of individualized medicine. Once developed, these
models are often used in diverse populations. Before being trusted for
clinical use in a new population, a prediction model needs to undergo
validation in a representative sample from that
population\textsuperscript{1}. For example, the original Framingham risk
scores for cardiovascular disease risk are developed using data from the
Framingham County in the US, but are externally validated for use in
many different populations\textsuperscript{2}.

During a validation study, the performance of the prediction model is
assessed in terms of average prediction error, discrimination (e.g.,
c-statistic), calibration (e.g., calibration intercept and slope), and
net benefit (NB)\textsuperscript{3}. Among such metrics, NB is a
decision-theoretic one, as it enables direct assessment of the clinical
utility of a model (i.e., if a model's expected NB is higher than that
of alternatives, it is expected to confer clinical utility). Due to such
decision-theoretic underpinnings, and despite its relatively new arrival
in the field of predictive analytics, NB has gained significant momentum
and has become a standard component of modern external validation
studies\textsuperscript{4}.

When interpreting the results of a validation study, the finite size of
the validation sample means that the assessment of model performance is
accompanied by uncertainty. Uncertainty in conventional metrics of model
performance is communicated using classical inferential methods (e.g.,
95\% confidence interval {[}CI{]} around c-statistic or calibration
slope). Similarly, sample size considerations when planning a validation
study are based on inferential statistics\textsuperscript{5,6}. However,
the relevance of this approach to uncertainty around NB as a
decision-theoretic metric is doubtful, as arbitrary significance levels
or pre-specified confidence bands do not necessarily translate to better
decisions\textsuperscript{7--9}.

Decision theory provides an alternative view to the consequences of
uncertainty by relating it to the outcome of decisions. A decision-maker
should adopt the strategy with the highest expected utility, which, if
implementation costs are comparable, is the one with the highest
expected NB. While uncertainty around NB should not affect the `adoption
decision', it is associated with utility loss as it hinders our ability
to identify the decision with the highest true NB. The extent of
uncertainty should thus inform the `research decision': whether the
expected utility loss is large enough to necessitate collecting further
evidence. This approach towards uncertainty quantification is referred
to as Value of Information (VoI) analysis\textsuperscript{10--12}.

VoI methods are widely accepted in health
policy-making\textsuperscript{11,13}. However, their relevance to
clinical decision-making and risk prediction has only recently been
highlighted\textsuperscript{14}. In particular, the expected value of
perfect information (EVPI), the expected gain in NB by completely
eliminating uncertainty, has been applied to both the development and
validation phases of risk prediction models\textsuperscript{14,15}.
However, to the best of our knowledge, the expected value of sample
information (EVSI), the expected gain in NB by conducting a study of a
given sample size, has not been defined and applied to clinical
prediction models.

The aim of the present work is to define EVSI for the external
validation studies of risk prediction models and propose algorithms for
its computation. The rest of this manuscript is structured as follows.
After outlining the context, we define EVSI for a future validation
study aimed at evaluating the clinical utility of a model in a new
population. We propose different algorithms for EVSI computations,
elaborate on their use case, and compare their performance in simulation
studies. A case study puts the developments in context. We conclude by
proposing areas of further inquiry.

\section{Methods}\label{methods}

\subsection{Context}\label{context}

We focus on validating a previously-developed risk prediction model in a
new target population. The developments are presented for prediction
models for binary responses but are generally applicable to any context
where NB can be assessed (e.g., survival outcomes\textsuperscript{16},
models for treatment benefit\textsuperscript{17}). A risk prediction
model is advertised as a function that maps patient characteristics to
an estimate of event risk. We have some `current' information about the
performance of this model in the target population (for example, from
expert opinion, or from a pilot study), but are uncertain if using the
model in this population is net beneficial. We are planning to obtain a
random sample \(\mathcal{D}\) containing \(\mathit{N}\) observations
from this population to reduce such uncertainty. This sample is expected
to improve our knowledge about model performance, which in turn will
increase our chance of making the correct decision whether to use the
model in this population. As such, procuring such a sample might be
associated with NB gain, and we are interested in evaluating the
expected gain in NB as a function of \(\mathit{N}\). We acknowledge the
relevance of other sources of uncertainty (e.g., missing data, the
plausibility that the sample is truly representative of the target
population) but focus on sampling uncertainty -- that is, our
uncertainty in true NBs because of the finite size of the validation
sample.

\subsection{Net benefit calculations for risk prediction
models}\label{net-benefit-calculations-for-risk-prediction-models}

Details of NB calculations for risk prediction models are explained in
its original paper and in many tutorials\textsuperscript{4,18,19}. In
brief, to turn a continuous predicted risk to a binary classification to
inform a treatment decision, a decision-maker needs to specify a risk
threshold \(z\) on predicted risks such that individuals classified as
high-risk are selected for the treatment (e.g., a 7.5\%-10\% threshold
on Framingham risk score is often used to initiate treatment with
statins to reduce cardiovascular disease risk\textsuperscript{20}).
Compared to treating no one (with a default utility of 0), the utility
of this classification can be expressed as \(p_{tp} - \omega p_{fp}\),
where \(p_{tp}\) and \(p_{fp}\) are the probabilities of true positive
and false positives, respectively, at this threshold, and
\(\omega \ge 0\) represents the utility tradeoff (exchange rate) between
a true positive and a false positive classification. Vickers and Elkin
showed that the value of \(\omega\) can be deduced from the risk
threshold itself\textsuperscript{18}: if a decision-maker is willing to
treat those with risk \(>z\) and not treat those with risk \(<z\), it
means among individuals with risk precisely equal to \(z\), they will be
ambivalent about the treatment decision. This ambivalence indicates that
in a population whose average outcome risk is \(z\), the expected
utility of treating everyone and treating no one is the same to the
decision-maker. Because in this population, treating everyone
corresponds to \(p_{tp}=z\) and \(p_{fp}=1-z\), we have
\(\omega=z/(1-z)\). For example, a 10\% threshold on cardiovascular risk
for initiating statin therapy means the benefit of treating one
individual who will experience a cardiovascular event is considered
equal to the harm of treating nine individuals who will not experience
such an event without treatment. This exchange rate enables calculating
NB in either net true positive or net false positive units for a given
threshold. In practice, the NB is calculated across a plausible range of
thresholds.

It is often more intuitive to express \(p_{tp}\) and \(p_{fp}\) in terms
of outcome prevalence and the model's true sensitivity and specificity
in the target population\textsuperscript{21}. Let \(\theta\) be the
triplet of outcome prevalence (\(\theta_p\), a quantity that is
independent of model performance), sensitivity (\(\theta_{se}\)), and
specificity (\(\theta_{sp}\)). We have \(p_{tp}=\theta_p\theta_{se}\)
and \(p_{fp}=(1-\theta_p)(1-\theta_{sp})\) (for brevity, we drop the
notation that indicates sensitivity and specificity are functions of
\(z\)).

Any given model should be compared with at least two `default'
strategies of treating no one and treating all. Treating no one has NB=0
by definition, and NB calculations for treating all follow the same
logic as above, but by considering all individuals as positive. Thus we
have three competing strategies: do not treat anyone, use the model to
treat those with predicted risk \(\ge z\), and treat all. We index them
by 0,1,2, respectively. This results in the following equation for NBs:

\begin{equation}
\label{master_NB_equation}
NB(i,\mathbf{\theta})=
\left\{
\begin{array}{lll}
0 & i=0 & \mbox{(treat no one)}
\\ 
\theta_p\theta_{se}-(1-\theta_p)(1-\theta_{sp})\frac{z}{1-z} & i=1 &(\mbox{use model to decide}) 
\\
\theta_p-(1-\theta_p)\frac{z}{1-z} & i=2 & (\mbox{treat all})
\\
\end{array}\right.,
\end{equation}

For time-to-event outcomes, \(\theta\)s and thus NBs are time-dependent
and should be defined at a time-horizon of interest\textsuperscript{16}.
If there are alternative strategies, they can also be considered (the
above indexing of NB() can accommodate other strategies), but this is
omitted here for the ease of exposition.

\subsection{Value of Information
analysis}\label{value-of-information-analysis}

VoI analysis is a Bayesian approach and requires explicit specification
of current information on the performance of the model in the target
population. By default, we assume this information is generated from a
preliminary validation exercise based on a random sample \(\mathbf{d}\)
of size \(n\). However, such information can be generated in other ways.
For example, one can extrapolate the performance of a model from its
development sample, or its performance in previous validation
studies\textsuperscript{22}, and accommodate any added uncertainty about
the compatibility of the populations by quantitatively discounting
external information\textsuperscript{23}. Formal expert elicitation
approaches can also be employed to construct distributions around model
performance metrics\textsuperscript{24}. In such instances,
\(\mathbf{d}\) can be an abstract entity representing, for example, the
expert-elicited information.

\subsection{The expected value of current
information}\label{the-expected-value-of-current-information}

Our current information about \(\theta\) based on the available evidence
is expressed as \(P(\theta | \mathbf{d})\). According to Bayes' theorem,
\(P(\theta|\mathbf{d}) \propto P(\theta)P(\mathbf{d}|\theta)\),
indicating that this information is influenced by any prior information
we might have had, as well as by \(P(\mathbf{d}|\theta)\), i.e., the
likelihood function, the support the observed data provide for a given
value of \(\theta\). With such current information, the decision-maker
should pick the strategy with the highest expected NB given the current
information:

\begin{equation}
{{E}} NB_\text{current}=\max_{_{i\in\{0,1,2\}}}{\mathbb{E}_{\mathbb{\theta}|\mathbf{d}} NB(i,\mathbb{\theta})},
\end{equation}

\noindent where the expectation is with respect to
\(P(\theta | \mathbf{d})\).

\subsection{The expected value of perfect information
(EVPI)}\label{the-expected-value-of-perfect-information-evpi}

Details of the reasoning and algorithms behind calculating EVPI are
presented previously\textsuperscript{15}. In brief, if we could know the
true values of \(\theta\), we would pick the strategy with the highest
true NB. We indeed do not have access to the ground truth, but have
partial information in terms of \(P(\theta | \mathbf{d})\). As such, we
can model the consequence of knowing the truth (having perfect
information) and take its expectation with respect to this distribution:

\begin{equation}
{{E}} NB_\text{truth}=\mathbb{E}_{\mathbf{\theta}|\mathbf{d}}\left[\max_{_{i\in\{0,1,2\}}} NB(i,\mathbf{\theta})\right].
\end{equation}

The EVPI is then the difference in expected NB of having perfect
information versus current information:

\begin{equation}
EVPI={{E}} NB_\text{truth}-{{E}} NB_\text{current}.
\end{equation}

The EVPI is a scalar quantity and quantifies the expected loss in NB due
to uncertainty, or alternatively, the expected gain in NB by completely
resolving uncertainty (i.e., conducting a validation study of infinite
size). We have proposed a generic algorithm based on bootstrapping to
estimate this expectation, as well as an asymptotic approach based on
central limit theorem and two-dimensional unit normal loss
integral\textsuperscript{15}.

\subsection{The expected value of sample information
(EVSI)}\label{the-expected-value-of-sample-information-evsi}

The reasoning behind EVSI calculation is similar to that of EVPI, with
the modification that instead of knowing the truth, we will have more
(but not perfect) information about \(\theta\), with additional
information coming from obtaining a future sample \(\mathcal{D}\). With
this added information we will have the opportunity to revise our
current decision. We will update our knowledge from
\(P(\theta | \mathbf{d})\) to \(P(\theta|\mathbf{d},\mathcal{D})\), and
will select the strategy that has the highest expected NB.

The NB of this approach once future data are obtained is
\(\max_i\mathbb{E}_\mathbf{\theta|{\mathcal{D},d}}NB(i,\mathbf{\theta})\).
We do not know what data we will observe in the future, but we can treat
\(\mathcal{D}\) as a random variable, and take the expectation of this
line of reasoning with regard to its predictive distribution (the
distribution for future observations given current data and the model):

\begin{equation}
{{E}} NB_\text{sample}=\mathbb{E}_{\mathcal{D} | \mathbf{d}}\left[ \max_{_{i\in\{0,1,2\}}}\mathbb{E}_{\mathbf{\theta}|\mathbf{d},\mathcal{D}}NB(i,\mathbb{\theta}) \right].
\end{equation}

The EVSI is the difference between this ENB and ENB under current
information:

\begin{equation}
EVSI={{E}} NB_\text{sample}-{{E}} NB_\text{current}.
\end{equation}

The EVSI is a non-negative scalar quantity in the same NB units as EVPI.
It quantifies the expected gain in NB by procuring a validation sample
of a given size. The higher the EVSI, the higher the expected gain from
the planned validation study. EVPI, quantifying the expected NB gain by
completely resolving uncertainty, puts an upper limit on EVSI.

\section{EVSI computation algorithms}\label{evsi-computation-algorithms}

The challenge for EVSI calculation is the two nested expectations in
\({{E}} NB_\text{sample}\). The inner expectation represents updated
estimates of NBs after obtaining future data, and the outer expectation
is with respect to the predictive distribution of future data. Below we
detail three computation algorithms.

\subsection{A bootstrap-based
algorithm}\label{a-bootstrap-based-algorithm}

If the current evidence \(\mathbf{d}\) is in terms of a random sample of
size \(n\) from the target population (e.g., from a pilot study), EVSI
computation can be performed via a two-level resampling algorithm. This
approach has been previously applied to EVSI computation for data-driven
economic evaluations\textsuperscript{25} and can be seen as a direct
extension of the bootstrap method for EVPI computation for external
validation studies\textsuperscript{15}. The logic behind this approach
is provided in detail previously\textsuperscript{25}. In a nutshell, let
\(\mathcal{D^*}\) be the target population from which \(\mathbf{d}\) and
\(\mathcal{D}\) are sampled. We take \(\mathcal{D^*}\) as a random
entity and apply the Bayes's rule:
\(P(\mathcal{D^*}|\mathbf{d})\propto P(\mathcal{D^*})P(\mathbf{d}|\mathcal{D^*})\).
Rubin showed that by using the improper prior
\(P(\mathcal{D^*})\sim \text{Dirichlet}(0,0,…,0)\) placed on all
possible observations in \(\mathcal{D^*}\), the posterior distribution
\(P(\mathcal{D^*}|\mathbf{d})\) will be a discrete distribution that
puts random weights \(\text{Dirichlet(1,1,…,1)}\) on each observation in
\(\mathbf{d}\). Since any parameter of interest \(\theta\) is a function
of \(\mathcal{D^*}\), posterior draws of weights from the
\(\text{Dirichlet}(1,1,…,1)\) distribution produce draws from the
posterior distribution of \(P(\theta|\mathbf{d})\)\textsuperscript{26}.
Further, once the generating population is at hand, the data of the
future study (\(\mathcal{D}\)) can be drawn by sampling from this
population. The resulting two-level resampling algorithm is explained in
Table \ref{table:BS_algorithm}.

\begin{center}
\begin{table}
  \noindent\fbox{%
      \parbox{\textwidth}{%
         \begin{enumerate}
    \item[-] \label{BS_curInfo} Prerequisite: Obtain $\mathbf{d}$, a pilot sample of size $n$ consisting of predicted risks and observed responses from the target population.\\
     \item For $j$ = 1 to $M$ (number of Monte Carlo simulations)\\
     \begin{enumerate}
       \item \label{BS_generate_population} Draw $\mathcal{D^*}$, a Bayesian bootstrap$^\dagger$ of the data $\mathbf{d}$, as a random draw from the posterior distribution of the population that has generated the sample. This is done by assigning random weights$^\ddagger$ $W^*\sim\text{Dirichlet}(1,1,...,1)$ to each observation in $\mathbf{d}$. \\
      \item \label{BS_trueThetas} Estimate $\theta^*$, the true values of prevalence, sensitivity, and specificity in this iteration, from $\mathcal{D^*}$. \\
      \item \label{BS_trueNBs} Calculate $NB^*_0$, $NB^*_1$, and $NB^*_2$, by plugging $\theta^*$ into Equation \ref{master_NB_equation}. These are true NBs in this iteration. Record the maximum of true NBs: $NB^*_\text{truth}=\max\{NB^*_0, NB^*_1, NB^*_2\}$. \\
      \item \label{BS_generate_data} Draw $\mathcal{D}$, a random realization of the data of the future study, through sampling $\mathit{N}$ observations with replacement from $\mathcal{D^*}$. \\
      \item \label {BS_futureSample} Create $\mathcal{D^{+}}= \mathbf{d} + \mathcal{D}$, the pooled data of existing and future samples. \\
      \item \label {BS_futureThetas} Calculate $\theta^{+}$ from  $\mathcal{D^{+}}$, the updated estimates of prevalence, sensitivity, and specificity after obtaining the future sample.  \\
      \item \label {BS_futureNBs} Calculate $NB^{+}_0$, $NB^{+}_1$, and $NB^{+}_2$, updated estimates of expected NBs after obtaining the future sample, by plugging $\theta^{+}$ into Equation \ref{master_NB_equation}. Determine the strategy that has the highest expected future NB: $l={\mbox{argmax}_{i \in \{0,1,2\}}(NB^{+}_i)}$. Record the true NB of this strategy: $NB^*_\text{sample}=NB^*_l$. \\
      \end{enumerate}
      \item[-] Next $j$ \\
      \item \label{curInfoNB} Take the average of $NB^*$s from step \ref{BS_trueNBs} and pick the maximum value. This is ${{E}} NB_{\text{current}}$$^\S$. \\
      \item Average $NB^*_\text{truth}$s from step \ref{BS_trueNBs}. This is ${{E}} NB_{\text{truth}}$. From this subtract ${{E}} NB_{\text{current}}$. This is EVPI. \\
      \item Average $NB^*_\text{sample}$s from step \ref{BS_futureNBs}. This is ${{E}} NB_{\text{sample}}$. From this subtract ${{E}} NB_{\text{current}}$. This is EVSI. 
   \end{enumerate}
      }%
  }
$^\dagger$ \small{Ordinary bootstrap can also be used - see text.}\\
$^\ddagger$ \small{One way of generating such wights is by generating $n$ exponential(1) random variables and scaling them to add up to 1.}\\
$^\S$ \small{Because the NB terms are linear on the elements of $\theta$, ${{E}} NB_{\text{current}}$ can also be directly evaluated in the original sample without bootstrapping, by plugging in the point estimates of $\theta$ into Equation \ref{master_NB_equation}. However, we recommend averaging over bootstrapped values as the positive correlation between this term and the other term in EVPI and EVSI equations improves precision and prevents getting occasional negative VoI estimates.}
\caption {Bootstrap-based computation of EVSI}
\label{table:BS_algorithm}
\end{table}
\end{center}

The power of this approach is in the flexibility of the bootstrap method
in accommodating different types of outcomes and practical
considerations in validation studies. For example, if \(\mathbf{d}\) has
non-trivial level of missing data, which affects the amount of
information it contains, Step \ref{BS_trueNBs} in Table
\ref{table:BS_algorithm} can include an imputation component, generating
the population \(\mathcal{D^*}\) without any missing values (and
similarly, Step \ref{BS_generate_data} can introduce missingness in the
future sample to simulate a real-world validation dataset).

The ordinary bootstrap, which assigns weights from a scaled
\(\text{Multinomial}(1,1,...,1)\) distribution to observations in
\(\mathbf{d}\), is conceptually similar to the Bayesian bootstrap and
has been interpreted in a Bayesian way (e.g., in the approximate
Bayesian bootstrap employed in missing value
imputation\textsuperscript{27}). In data-driven VoI analyses it
generates similar results to the Bayesian bootstrap\textsuperscript{25}.
It thus can be used in lieu of the Bayesian bootstrap in step
\ref{BS_generate_population}.

\subsection{A fast algorithm based on beta-binomial modeling for binary
outcomes}\label{a-fast-algorithm-based-on-beta-binomial-modeling-for-binary-outcomes}

In their proposal for a parametric Bayesian NB estimation for binary
outcomes, Cruz and Korthauer note that the likelihood function for
\(NB\) can be factorized into terms involving prevalence, sensitivity,
and specificity\textsuperscript{28}. As such, they take advantage of the
beta-binomial conjugacy and note that specifying prior information in
terms of independent beta distributions on these quantities will result
in corresponding posterior distributions that are also independent beta.
In this case, given that \(P(\theta|\mathbf{d})\) is composed of three
independent beta distributions, by the same token
\(P(\theta|\mathbf{d},\mathcal{D})\) after a realization of
\(\mathcal{D}\) will also be composed of three independent beta
distributions. This results in a fast Monte Carlo approach for EVSI
computation as explained in Table \ref{table:betabin_algorithm}.

\begin{center}
\begin{table}
\noindent\fbox{%
\parbox{\textwidth}{%
\begin{enumerate}
  \setcounter{enumi}{0}
  \item[-] \label{betabin_curInfo} Prerequisite: Specify current information $P(\theta | \mathbf{d})$ as\\
      $\theta_p \sim \text{Beta}(\alpha_{p},\beta_{p})$, 
      $\theta_{se} \sim \text{Beta}(\alpha_{se}, \beta_{se})$,
      $\theta_{sp} \sim \text{Beta}(\alpha_{sp}, \beta_{sp})$. \\
  \item For $j$ = 1 to $M$ (number of Monte Carlo simulations) \\
  \begin{enumerate}
    \item \label{betabin_trueThetas} Obtain $\theta^*$, by random drawing from $P(\theta | \mathbf{d})$. These are true $\theta$s in this iteration. \\
    \item \label{betabin_trueNBs} Calculate $NB^*_0$, $NB^*_1$, and $NB^*_2$, by plugging $\theta^*$ into Equation \ref{master_NB_equation}. These are true NBs in this iteration. Record the maximum true NB: $NB^*_\text{truth}=\max\{NB^*_0, NB^*_1, NB^*_2\}$. \\
    \item \label{betabin_futureSample} Generate $\mathcal{D}$, the sample of future study, defined by $\{\mathit{N}_{tp}, \mathit{N}_{fn}, \mathit{N}_{tn}, \mathit{N}_{fp}\}$ given $\theta^*$ obtained in Step \ref{betabin_trueThetas} as:\\
      $\mathit{N}_{+}\sim \text{Binomial}(N,\theta_p^*)$ (number of positive cases in the future sample), \\
      $\mathit{N}_{tp}\sim \text{Binomial}(\mathit{N}_{+},\theta_{se}^*)$,\\
      $\mathit{N}_{fn}=\mathit{N}_{+}-\mathit{N}_{tp}$,\\
      $\mathit{N}_{tn}\sim \text{Binomial}(\mathit{N}-\mathit{N}_{+},\theta_{sp}^*)$,\\
      $\mathit{N}_{fp}=\mathit{N}-\mathit{N}_{+}-\mathit{N}_{tn}$. \\
    \item\label{betabin_futureThetas} Calculate $\theta^{+}$, the updated estimates of prevalence, sensitivity, and specificity after obtaining the future sample. :\\
      ${\theta^{+}_p}=(\alpha_{p}+\mathit{N}_{tp}+\mathit{N}_{fn})/(\alpha_{p}+\beta_{p}+\mathit{N})$, \\
      ${\theta^{+}_{se}}=(\alpha_{se}+\mathit{N}_{tp})/(\alpha_{se}+\beta_{se}+\mathit{N}_{tp}+\mathit{N}_{fn})$,\\
      ${\theta^{+}_{sp}}=(\alpha_{sp}+\mathit{N}_{tn})/(\alpha_{sp}+\beta_{sp}+\mathit{N}_{tn}+\mathit{N}_{fp})$. \\
    \item \label{betabin_futureNBs} Calculate $NB^{+}_0$, $NB^{+}_1$, and $NB^{+}_2$, updated estimates of NBs after obtaining the future sample, by plugging $\theta^{+}$ into Equation \ref{master_NB_equation}. Determine the strategy that has the highest expected future NB: $l={\mbox{argmax}_{i \in \{0,1,2\}}(NB^{+}_i)}$. Record the true NB of this strategy: $NB^*_\text{sample}=NB^*_l$. \\
    \end{enumerate}
  \item[-] Next $j$ \\
  \item \label{betabin_curInfoNB} Take the average of $NB^*$s from step \ref{betabin_trueNBs} and pick the maximum value. This is ${{E}} NB_{\text{current}}$. \\
  \item Average $NB^*_\text{truth}$s from step \ref{betabin_trueNBs}. This is ${{E}} NB_{\text{truth}}$. From this subtract ${{E}} NB_{\text{current}}$. This is EVPI. \\
  \item Average $NB^*_\text{sample}$s from step \ref{betabin_futureNBs}. This is ${{E}} NB_{\text{sample}}$. From this subtract ${{E}} NB_{\text{current}}$. This is EVSI. \\
  \end{enumerate}
}%
}
\caption {Computation of EVSI based on beta-binomial modeling}
\label{table:betabin_algorithm}
\end{table}
\end{center}

Independent beta distributions for each component of \(\theta\) can
emerge, for example, if we ask an expert to express their belief about
prevalence, sensitivity, and specificity, in terms of fractions. That
is, if their best estimate of prevalence is 30\%, and their uncertainty
is akin to their opinion comping from a sample of 10 individuals, the
information can be expressed as \(\theta_p \sim \text{Beta}(3,7)\)
(similar for sensitivity and specificity).

Importantly, for binary outcomes and in the absence of missing data, the
Bayesian bootstrap approach explained previously reduces to this
beta-binomial model. This is because the information in a sample
\(\mathbf{d}\) for estimating NBs can be fully specified via four
quantities: the number of true positives (\(n_{tp}\)), false negatives
(\(n_{fn}\)), true negatives (\(n_{tn}\)), and false positives
(\(n_{fp})\). Due to the aggregate property of the Dirichlet
distribution\textsuperscript{29}, the \(\text{Dirichlet}(0,0,...,0)\)
prior used in the Bayesian bootstrap is equal to assigning
\(\text{Beta}(0,0)\) priors to \(\theta_p\), \(\theta_{se}\), and
\(\theta_{sp}\), resulting in the posterior distribution for the
population that can be specified as
\((p_{tp}, p_{fn}, p_{tn}, p_{fp}) \sim \text{Dirichlet}(n_{tp}, n_{fn}, n_{tn}, n_{fp})\).
As such, the beta-binomial algorithm will be equal to the Bayesian
bootstrap and can be used instead for VoI computations due to its
computational efficiency.

Yet another utility of this approach is that instead of the improper
prior employed in the Bayesian bootstrap, the user can specify
informative priors. Imagine in the current sample \(\mathbf{d}\) of size
\(n\), there are \(m\) individuals who have experienced the outcome. If
an independent cross-sectional study in the same target population has
estimated outcome prevalence that can be expressed as
\(\text{Beta}(\alpha, \beta)\), our current information about prevalence
can be expressed as \(\theta_p \sim \text{Beta}(\alpha+m, \beta+n-m)\).
Even in the absence of external information, weakly informative priors
can be employed to rectify some of the previously-identified problems
with bootstrap-based VoI calculations at extreme situations. In small
validation samples and extreme thresholds, VoI calculations might
generate counter-intuitive results\textsuperscript{15}. One instance is
when sample estimates of \(\theta_{se}\) or \(\theta_{sp}\) are at their
extreme (0 or 1). The improper \(\text{Beta}(0,0)\) prior on these
quantities results in improper beta posteriors with one of their two
parameters being 0. This can be resolved if one uses a flat
\(\text{Beta}(1,1)\) prior for these quantities.

\subsection{\texorpdfstring{A general algorithm for arbitrary
specification of
\(P(\theta | \mathbf{d})\)}{A general algorithm for arbitrary specification of P(\textbackslash theta \textbar{} \textbackslash mathbf\{d\})}}\label{a-general-algorithm-for-arbitrary-specification-of-ptheta-mathbfd}

In many practical situations our current information
\(P(\theta|\mathbf{d})\) may not have a simple mathematical form, but
one can still draw a Monte Carlo sample of arbitrary size from its
distribution. Examples include when \(P(\theta_p)\) is constructed from
a meta-analysis of prevalence studies, or \(P(\theta_{se},\theta_{sp})\)
from a bivariate meta-analysis of the diagnostic accuracy of the test,
giving rise to a joint distribution for
\(P(\theta_{se},\theta_{sp})\)\textsuperscript{30}. Markov-Chain
Monte-Carlo methods can be used to obtain samples from the joint
distribution of \(\theta\), even though the mathematical form of the
joint density might be intractable (an example is the case studies in
Wynants et al\textsuperscript{21}).

For such a general case where we have a sample from
\(P(\theta|\mathbf{d})\), the empirical distribution of the sample can
be used as the distribution for the current information. We note that
\(P(\theta|\mathbf{d},\mathcal{D})\propto P(\theta)P(\mathbf{d},\mathcal{D}|\theta) =P(\theta)P(\mathbf{d}|\theta)P(\mathcal{D}|\theta) \propto P(\theta|\mathbf{d})P(\mathcal{D}|\theta)\).
As such, the posterior distribution of \(\theta\) after observing the
future sample can be approximated by the discrete distribution that puts
a mass \(w^{(i)} \propto P(\mathcal D|\theta^{(i)})\) on the \(i\)th
observation of the sample from \(P(\theta | \mathbf{d})\). This will
result in the general algorithm explained in Table
\ref{table:general_algorithm}.

\begin{center}
\begin{table}
\noindent\fbox{%
  \parbox{\textwidth}{%
    \begin{enumerate}
      \item[-]\label{general_empiricalDist} Prerequisite: Specify current information as arbitrary distribution $P(\theta | \mathbf{d})$. Obtain $M$ draws $(\theta^{(1)},\theta^{(2)},...,\theta^{(M)})$ from $P(\theta | \mathbf{d})$. \\
      \item For $j$ = 1 to $M$ (number of Monte Carlo simulations) \\
      \begin{enumerate}
        \item \label{general_trueThetas} Let $\theta^*=\theta^{(i)}$. This is true $\theta$ in this iteration$\dagger$. \\
        \item \label{general_trueNBs} Calculate $NB^*_0$, $NB^*_1$, and $NB^*_2$, by plugging $\theta^*$ into Equation \ref{master_NB_equation}.  These are true NBs in this iteration. Record the maximum true NB: $NB^*_\text{truth}=\max\{NB^*_0, NB^*_1, NB^*_2\}$. \\
        \item \label{general_futureSample} Generate $\mathcal{D}$, the sample of future study, defined by $\{\mathit{N}_{tp}, \mathit{N}_{fn}, \mathit{N}_{tn}, \mathit{N}_{fp}\}$ as follows:\\ 
        $\mathit{N}_{+}\sim \text{Binomial}(\mathit{N},\theta^*_p)$ (number of positive cases in the future sample), \\
        $\mathit{N}_{tp}\sim \text{Binomial}(\mathit{N}_{+},\theta^*_{se})$,\\
        $\mathit{N}_{fn}=\mathit{N}_{+}-\mathit{N}_{tp}$,\\
        $\mathit{N}_{tn}\sim \text{Binomial}(\mathit{N}-\mathit{N}_{+},\theta^*_{sp})$,\\
        $\mathit{N}_{fp}=\mathit{N}-\mathit{N}_{+}-\mathit{N}_{tn}$. \\
        \item \label {general_futureThetas} Update the distribution of $\theta$ given the future sample. This is done by updating  the weights $w^{(k)}$ based on the likelihood $P(\mathcal{D}| \theta)$. For binary outcomes the weights are:\\
        $w^{(k)} \propto P_{Bin}(\mathit{N}_{tp}+\mathit{N}_{fn},\mathit{N},\theta_{p}^{(k)}) \times P_{Bin}(\mathit{N}_{tp},\mathit{N}_{tp}+\mathit{N}_{fn},\theta_{se}^{(k)}) \times P_{Bin}(\mathit{N}_{tn},\mathit{N}_{tn}+\mathit{N}_{fp},\theta_{sp}^{(k)})$,\\
        with $P_{Bin}(x,m,p)$ being the probability mass function of the binomial distribution at value $x$ for $m$ trials and success probability  $p$. \\
        \item \label{general_futureNBs} Calculate $NB^{+}_0$, $NB^{+}_1$, and $NB^{+}_2$, updated estimates of expected NBs after obtaining the future sample, by using the weights generated in the previous step as follows:\\
        $NB^{+}_i=
          \left\{
          \begin{array}{ll}
          0 & i=0 
          \\ 
          \sum_{k=1}^M w^{(k)}\{\theta^{(k)}_p\theta^{(k)}_{se}-(1-\theta^{(k)}_p)(1-\theta^{(k)}_{sp})\frac{z}{1-z}\}/\sum_{k=1}^M w^{(k)} & i=1 
          \\
          \sum_{k=1}^M w^{(k)}\{\theta^{(k)}_p-(1-\theta^{(k)}_p)\frac{z}{1-z}\}/\sum_{k=1}^M w^{(k)} & i=2 
          \\
          \end{array}\right.. $
      
      Determine the strategy that has the highest expected NB in this pooled sample: $l={\mbox{argmax}_{i \in \{0,1,2\}}(NB^{+}_i)}$. Record the true NB of this strategy: $NB^*_\text{sample}=NB^*_l$. \\
      \end{enumerate}
      \item[-] Next $j$ \\
      \item Average $NB^*_\text{truth}$s from from step \ref{general_trueNBs}. This is ${{E}} NB_{\text{truth}}$. Subtract from this ${{E}} NB_{\text{current}}$. This is EVPI. \\
      \item Average $NB^*_\text{sample}$s from step \ref{general_futureNBs}. This is ${{E}} NB_{\text{sample}}$. Subtract from this ${{E}} NB_{\text{current}}$. This is EVSI. \\
    \end{enumerate}
  }%
}
$\dagger$ \small {If the size of the sample is large it might not be computationally feasible to loop over all observations. In which case one can pick one of $\theta^{(i)}s$ at random in this step, but still uses the entire sample in step \ref{general_futureNBs}.}
\caption{EVSI computation for binary outcomes for general, non-conjugate distributions}
\label{table:general_algorithm}
\end{table}
\end{center}

\subsection{Case Study}\label{case-study}

We used data from GUSTO-I, a clinical trial of multiple thrombolytic
strategies for acute myocardial infarction (AMI), as our case
study\textsuperscript{31}. This dataset has been widely used for
methodological research in predictive analytics\textsuperscript{32--34}.
All analyses are conducted in R\textsuperscript{35}, with a fast
implementation of the general EVSI algorithm in C++ (please refer to the
\emph{evsiexval} R package for details).

For the present case study, we use a similar setup as our previous work,
with a risk prediction model for 30-day mortality developed using the
non-US sample of this dataset, and validated in the US sub-sample. The
risk prediction model is a logistic regression model fitted to the
entire non-US subset (sample size 17,796). We initially assume we have
access to only \(n=500\) individuals from the US sub-sample as the
source of current information. We randomly selected 500 individuals,
without replacement, from the entire 23,034 observations in the US
sub-sample of GUSTO-I. Later, we will use more observations from the US
sub-sample to study how EVSI behaves with the amount of current
information. We consider the range of relevant thresholds to be
1\%-10\%, with 1\%, 2\%, 5\%, and 10\% being of primary interest. VoI
computations are based on 10\^{}6 Monte Carlo simulations using the
aggregate beta-binomial model, with a flat \(\text{Beta}(1,1)\) prior on
each component of \(\theta\).

Using this setup, we performed two simulation studies to explore the
face validity of the proposed algorithms. In the first set, we aimed at
comparing the numerical stability and computational time of the three
algorithms. In the second set, we explored how EVSI changes as a
function of the amount of current information, represented by \(n\), the
size of the current validation data from which \(P(\theta|\mathbf{d})\)
is constructed.

\section{Results}\label{results}

The prevalence of the outcome in the development sample (entire non-US
sub-sample) was 0.0723; in the current validation sample it was 0.0860,
while in the entire US sub-sample it was 0.0679. The risk prediction
model was of the form:

logit(P(Y=1))= - 2.0842 + 0.0781{[}age{]} + 0.4027{[}AMI location other
(vs.~inferior){]} + 0.5773{[}AMI location anterior (v. inferior){]} +
0.4678{[}previous AMI{]} + 0.7666{[}AMI severity (Killip score){]} -
0.0775{[}min(blood pressure, 100){]} + 0.0182{[}pulse{]}.

\begin{figure}
\centering
  \begin{tabular}{@{}c@{}}
    \includegraphics [width=1.0\linewidth,height=160pt]{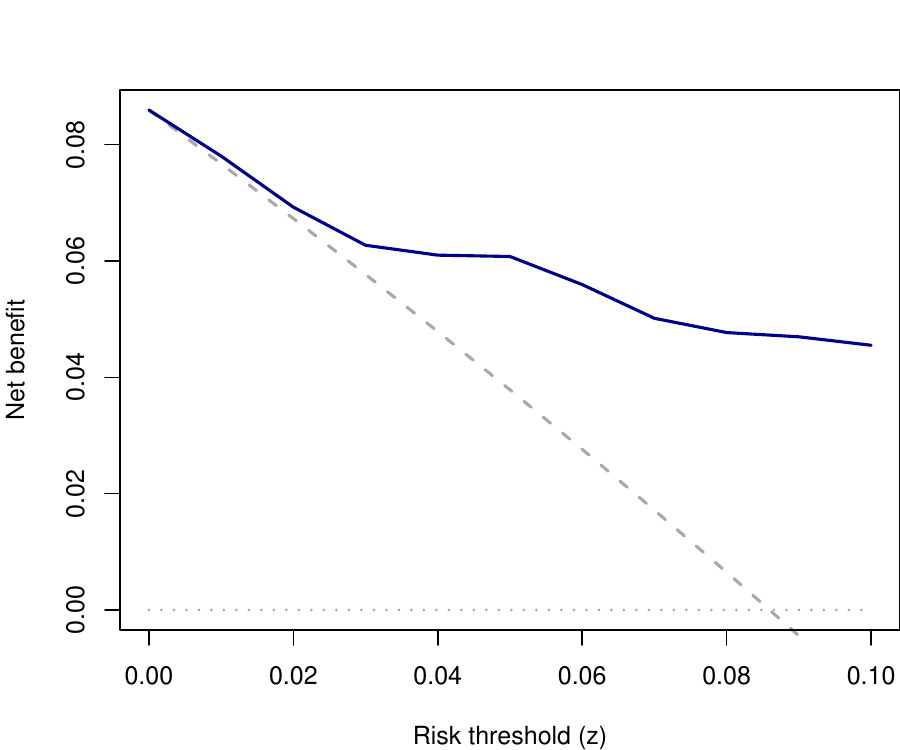} \\ [\abovecaptionskip]
    \small  (a) Decision curve. Solid blue: use the model; dashed gray: treat all; dotted gray: treat none
  \end{tabular}
  \begin{tabular}{@{}c@{}}
    \includegraphics [width=1.0\linewidth,height=160pt]{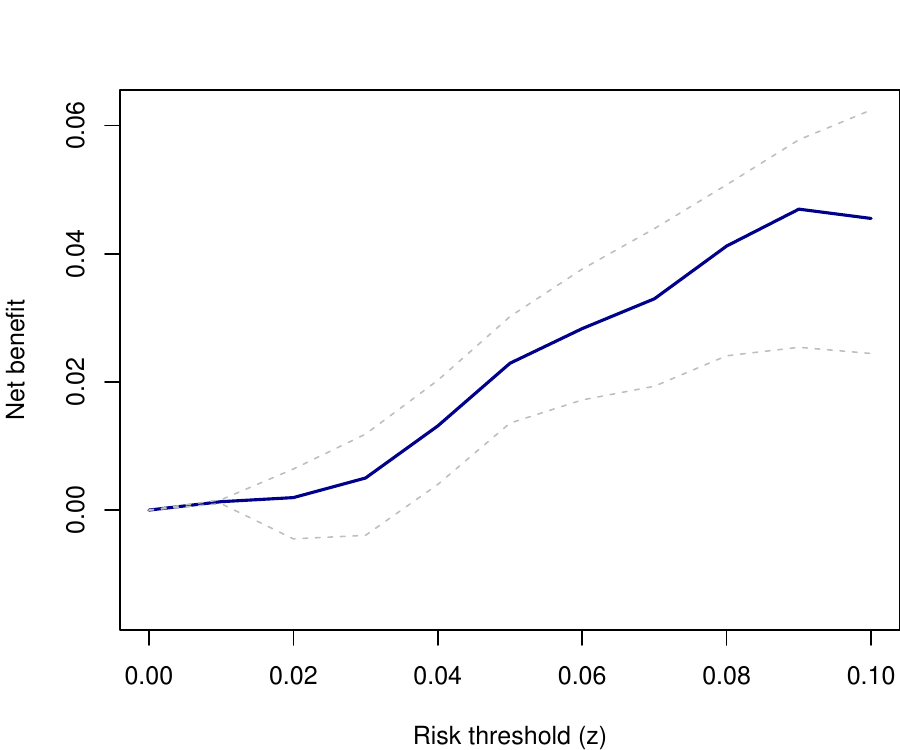} \\ [\abovecaptionskip]
    \small (b) Incremental NB of the model versus best alternative strategy
  \end{tabular}
\caption{Decision curve (a) and incremental net benefit of the model versus the best alternative strategy (b). Dashed lines are 95\% confidence intervals}
\label{fig:GUSTO_DCA_dNB}
\end{figure}

The decision curve depicting the NB of the model alongside alternative
strategies is presented in Figure \ref{fig:GUSTO_DCA_dNB}. Panel (a)
shows the NB, and panel (b) shows the difference between the NB of the
model and the best alternative default strategy
(\(NB_1-\max(NB_0,NB_2)\)) along with its 95\% CI (based on the
percentile methods using 10,000 bootstraps). The model had higher
expected NB than the alternative strategies at the entire range of
thresholds of interest.

Among 500 observations in the current sample, 43 experienced the
outcome. Given the prior \(\text{Beta}(1,1)\), this will result in
\(\theta_p\sim \text{Beta}(44,458)\). The distribution of sensitivity
and specificity depends on the threshold. Taking \(z=0.02\) as an
example, given that 41 of the 43 individuals who experienced the outcome
had predicted risk \(\ge z\), and 147 of the 457 who did not experience
the outcome had predicted risk \(<z\), with a \(\text{Beta}(1,1)\)
prior, we have \(\theta_{se}\sim\text{Beta}(42, 3)\) and
\(\theta_{se}\sim\text{Beta}(148, 311)\).

\begin{figure}
\centering
  \includegraphics [width=1.0\linewidth,height=200pt]{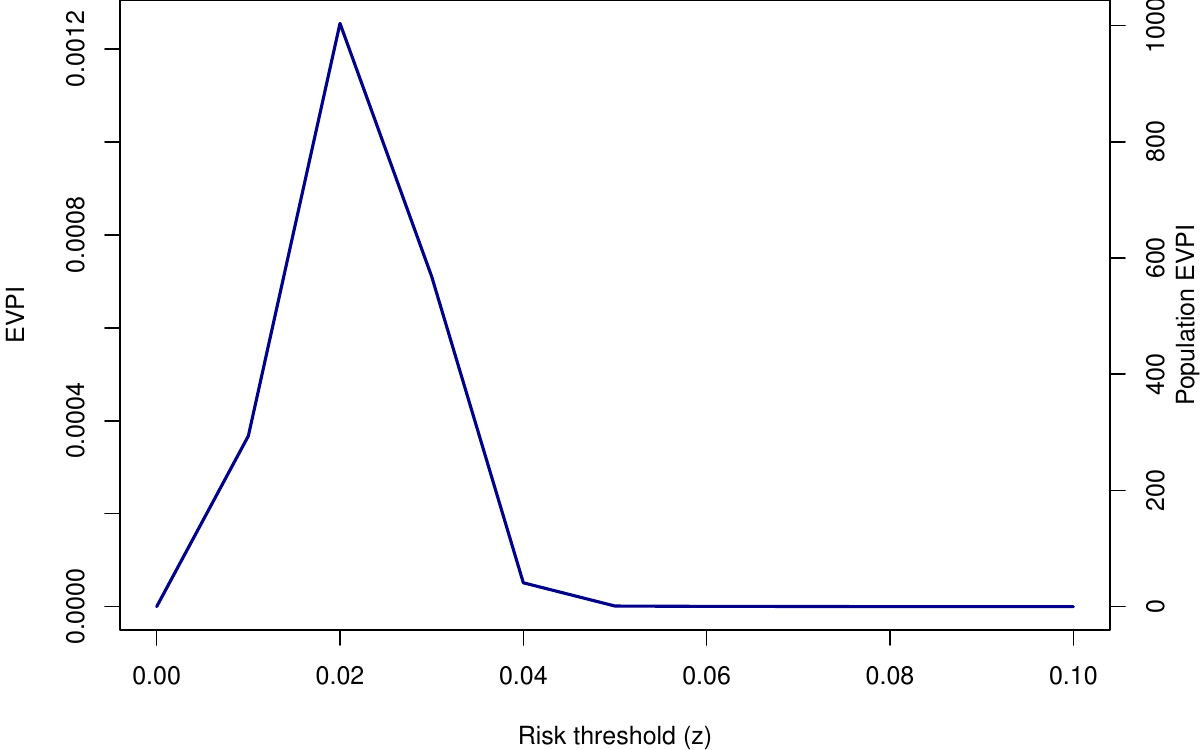} \\ [\abovecaptionskip]
\caption{The expected value of perfect information (EVPI) of the validation sample}
\label{fig:GUSTO_EVPI}
\end{figure}

The EVPI curve is presented in Figure \ref{fig:GUSTO_EVPI}. The EVPI at
0.01 threshold was 0.00037. At the 0.02 threshold it was 0.00125. The
EVPI was 0 at 0.05 and 0.10 thresholds (so EVSI for any \(\mathit{N}\)
will also be 0). We therefore focus on VoI calculations at 0.01 and 0.02
thresholds. EVPI quantifies the expected NB loss per decision (every
time the model might be used). The overall NB loss due to uncertainty is
therefore affected by the expected number of times the medical decision
will be made\textsuperscript{36}. In scaling VoI quantities to the
population, we apply a similar approach as in our previous
work\textsuperscript{15}: There are around 800,000 AMIs in the US every
year\textsuperscript{37}, and a national guideline development panel in
charge of recommending risk stratification for AMI management can
consider all those instances relevant. The scaled VoI values are
provided in true positive units on the second Y-axis of the EVPI figure.
At the 0.02 threshold, a future validation study can have a maximum
benefit equal to gaining 1,004 net true positive (individuals who will
die within 30 days and are correctly identified as high risk) per year,
or avoiding 49,188 extra false positive cases per year (individuals who
will not die within 30 days but are identified as high risk).

\begin{figure}
\centering
  \begin{tabular}{@{}c@{}}
    \includegraphics [width=2.0\linewidth,height=200pt]{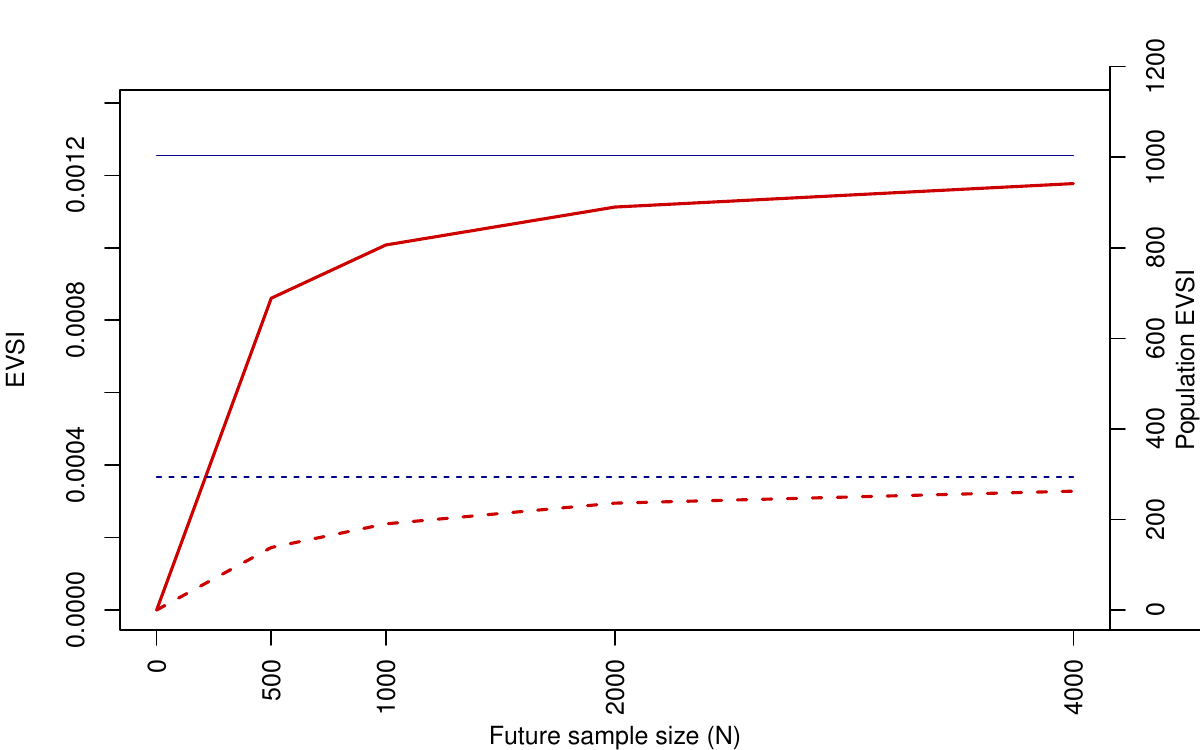} \\ [\abovecaptionskip]
  \end{tabular}
\caption{The expected value of sample information (EVSI) for the case study. Red: EVSI; blue (horizontal): the expected value of perfect information (EVPI). Dashed lines: $z$ (risk threshold)=0.01; solid lines: $z$=0.02}
\label{fig:GUSTO_EVSI}
\end{figure}

The EVSI curves are shown in Figure \ref{fig:GUSTO_EVSI}. Generally, it
is expected that EVSI will increase with a larger future study, and will
asymptote to EVPI, a pattern that is obvious for both thresholds. The
`diminishing return' pattern is also clear: the expected NB gain steeply
rises at small samples and plateaus as \(\mathit{N}\) grows, leaving
little room for gaining NB beyond \(\mathit{N}\)=4,000. Similar to EVPI,
population scaling is applied to EVSIs and is presented as the second
Y-axis of the EVSI curve. At 0.02 threshold, a future study of size
\(\mathit{N}=1,000\) has a per-decision EVSI value of 0.00101,
corresponding to population value of 806 in true positive cases gained,
or 39,500 in false positive cases averted.

\section{Discussion}\label{discussion}

Evaluating the performance of a model in a finite sample is fraught with
uncertainty. In this work, we applied the VoI framework to investigate
the decision-theoretic consequences of such uncertainty. We defined
validation EVSI as the expected gain in NB by obtaining an external
validation sample of a given size from the target population of
interest. In a case study we showed the feasibility of EVSI
calculations, and studied how EVSI is affected by the amount of current
information and the sample size of the future study. We suggested
scaling the EVSI to the population to quantify the overall gain in
clinical utility in true (or false) positive units from conducting an
external validation study of a given sample size.

We proposed three algorithms for EVSI computations (with implementation
in R as part of the \emph{evsiexval} package:
\url{https://github.com/resplab/evsiexval}). The bootstrap-based
algorithm is applicable when previous individual-level data (e.g., from
a pilot validation study) are at hand. The main advantage of this
algorithm is in its flexibility, for example in mimicking expected
patterns of missingness in data. Among the three algorithms, this
algorithm is the one that can most readily be extended to survival
outcomes (similar to the extension of the bootstrap method for inference
around NB\textsuperscript{16}). However, this algorithm is applicable
only when individual-level data are available and can be slow. The
beta-binomial algorithm is applicable when current evidence on model
performance can be expressed as independent beta distributions for
prevalence, sensitivity, and specificity, and is the fastest of the
three algorithms. We showed that for binary outcomes and in the absence
of censoring this algorithm is equal to the Bayesian bootstrap (with the
added flexibility that prior information can be incorporated). The
general algorithm is the most versatile one as it works with any joint
distribution of \(\theta\) as long as one can obtain random draws from
it. As such, this algorithm can be useful when
\(P(\theta | \mathbf{d})\) does not have a tractable form. However, the
finite size of the sample adds another layer of uncertainty, potentially
requiring significantly more computation time to achieve the same
numerical accuracy of the other two algorithms.

How can EVSI analysis inform study design in predictive analytics?
Conducting a validation study is an investment in resources that will
generate further information on the performance of a clinical prediction
model in a target population. The decision whether to undertake a
validation study should ultimately hinge on whether the information
gained from such a study is worth the required investments. The EVSI,
when scaled to the population, determines the expected return on
investment in NB unit. Ultimately, such return should be contrasted
against the efforts and resources required for such a study. In
decision-analytic (health policy) modeling, EVSI is typically in net
monetary units, and when scaled to population, can be compared with the
budget of a planned data collection activity. The optimal sample size
will be one that maximizes the difference between population EVSI and
study costs\textsuperscript{36}. The NB for risk prediction models, on
the other hand, is in net true or false positive units, and as such
cannot directly be compared with the budget of a validation study. One
can always embark on full decision analysis to translate all outcomes to
net monetary units, but this will likely require sophisticated decision
modeling and context-specific assumptions on long-term outcomes, a
process that might take significant amount of time and require further
data collection (e.g., to obtain utility weights for outcomes). To us, a
main reason for the vast popularity of decision curve analysis is that
it provides an assumption-free, reproducible method for NB calculation
based on the very same data that are used for studying model calibration
and discrimination. We think VoI analysis during model development and
validation should generally keep the same spirit. A full decision
analysis should be relegated to after an impact analysis has measured
the resource-use implications of implementing the model. This, however,
means the decision rules for determining the optimal sample size of
development and validation studies based on the VoI framework would be
different than those used in health policy analysis. This paper
deliberately stayed away from proposing such rules, and instead focused
on defining concepts and proposing computation methods for EVSI.
Proposing such decision rules is detached from EVSI calculations and
deserves its own airing.

There are multiple areas of further inquiry. The EVSI framework should
also be applied to the development phase of prediction models. This can
guide the investigator on whether further development, or moving to
validation, should be prioritized. We mainly focused on NB loss due to
sampling uncertainty. However, there are several sources of uncertainty,
such as whether our existing information on the performance of the model
is directly applicable to the target population, or if predictors and
outcome are measured with the same quality between the study and usual
practice. The comparative statistical and computational performance of
the EVSI computations algorithms, and the adequacy of a given number of
simulations for each algorithm, should be evaluated in dedicated
studies. We also do not claim the algorithms we proposed are the only
ones that can be used for EVSI computations. Other algorithms, such as
those based on central limit theorem\textsuperscript{38,39}, can prove
useful. VoI analysis in decision modeling has received significant boost
in computational speed in recent years due to the arrival of algorithms
based on non-parametric regression modeling\textsuperscript{40,41}. This
approach can facilitate VoI analysis in risk prediction as well. The
EVSI defined in this work is for a single, homogeneous target
population, and does not consider heterogeneous settings. VoI metrics
and corresponding computation algorithms for multi-center studies should
be developed separately. Further, during external validation, often a
secondary aim is to update the model if its performance turns out to be
sub-optimal\textsuperscript{42}. Such model revision can take different
levels of complexity\textsuperscript{43} and one might be interested in
the expected yield of a given sample size for such model revision. This
will get connected to the VoI concepts for model development. As stated
earlier, how VoI metrics for prediction models should inform objective
functions for determining the optimal sample size should be debated by
the community in the hope of generating consensus and best practice
standards.

There is an ongoing debate on the appropriateness of conventional
metrics of uncertainty when reporting NB of a clinical prediction
model\textsuperscript{7--9}. The same concerns can logically be extended
to frequentist method for sample size and power calculations around
NB\textsuperscript{5,6}. VoI methodology provides a rigorous,
utilitarian response to such controversies. The toolbox of VoI methods
for clinical prediction models is growing, and perhaps it is time to
formalize the role of VoI in uncertainty quantification and design of
empirical studies in predictive analytics.

\newpage

\section*{References}\label{references}
\addcontentsline{toc}{section}{References}

\phantomsection\label{refs}

\end{document}